\documentclass[preprintnumbers]{revtex4}
\usepackage{amsfonts}
\usepackage{amsmath}
\usepackage{hyperref}
\usepackage{amssymb}
\usepackage{subfigure}
\usepackage[english]{babel}
\usepackage{graphicx}
\usepackage{epsfig}
\usepackage{bm}
\usepackage{color}
\usepackage{longtable}
\usepackage{verbatim}
\usepackage{longtable}
\usepackage[utf8]{inputenc}

\newcommand{\mathsym}[1]{{}}

\input{tcilatex}
\begin{document}

\title{Phenomenology of an extended IDM with loop-generated fermion mass
hierarchies}
\author{A. E. C\'arcamo Hern\'andez$^{{a}}$}
\email{antonio.carcamo@usm.cl}
\author{Sergey Kovalenko$^{{a}}$}
\email{sergey.kovalenko@usm.cl}
\author{Roman Pasechnik{}$^{b,c,d}$}
\email{Roman.Pasechnik@thep.lu.se}
\author{Ivan Schmidt$^{{a}}$}
\email{ivan.schmidt@usm.cl}
\affiliation{$^{{a}}$Universidad T\'ecnica Federico Santa Mar\'{\i}a and Centro Cient%
\'{\i}fico-Tecnol\'ogico de Valpara\'{\i}so\\
Casilla 110-V, Valpara\'{\i}so, Chile\\
$^{{b}}$Department of Astronomy and Theoretical Physics, Lund University,
Solvegatan 14A, SE-223 62 Lund, Sweden\\
$^{{c}}$Nuclear Physics Institute ASCR, 25068 \v{R}e\v{z}, Czech Republic\\
$^{{d}}$Departamento de F\'isica, CFM, Universidade Federal de Santa
Catarina, C.P. 476, CEP 88.040-900, Florian\'opolis, SC, Brazil }

\begin{abstract}
We perform a comprehensive analysis of the most distinctive and important
phenomenological implications of the recently proposed mechanism of sequential loop generation of strong hierarchies in
the Standard Model (SM) fermion mass spectra. This mechanism is consistently
realized at the level of renormalizable interactions in an extended variant
of the Inert Higgs Doublet model, possessing the additional $Z_{2}^{(1)}\times Z_{2}^{(2)}$ discrete and $U_{1X}$ gauge family
symmetries, while the matter sectors of the SM are extended by means of $SU_{2L}$-singlet scalars, heavy vector-like leptons and quarks, as well as right-handed neutrinos. We thoroughly analyze the most stringent constraints
on the model parameter space, coming from the $Z^{\prime }$ collider
searches, related to the anomaly in lepton universality, and the muon
anomalous magnetic moment, as well as provide benchmark points for further
tests of the model and discuss possible ``standard candle'' signatures
relevant for future explorations.
\end{abstract}

\maketitle



\section{Introduction}

\label{Sec:Intro} 

The hypothetical extensions of the Standard Model (SM) that accommodate a
dynamical explanation of the mass and mixing hierarchies in the quark,
lepton and neutrino sectors, are typically expected to contain many new
interactions and states at high scales of the theory. In particular,
additional scalar fields are required to break the high-scale (e.g. discrete
or continuous family) symmetries, causing the formation of specific patterns
in the fermion mass spectra across generations. The additional inert
sectors, such as heavy right-handed neutrinos, are mandatory for see-saw
type mechanisms of neutrino mass generation, and play a supplemental but
important cosmological role in leptogenesis and also as candidates for DM.
In practice, there are no strong constraints on how many additional heavy
scalar singlet and vector-like fermion states could be added to the SM at
the fundamental level, as they typically produce vanishing direct signatures
in collider measurements, but may have indirect (e.g. via radiative
corrections) signatures imprinted into the patterns of SM couplings and mass
parameters.

In general, additional states are required to explain specific patterns in
the SM fermion spectra. For example, to address only the quark sector and to
explain the Cabbibo-like structure of the quark mixing simultaneously with
the hierarchies in the quark mass spectrum, the addition of a gauged $U_{1X}$
or discrete family symmetry and few extra scalar fields seems to be enough
(see Refs.~\cite%
{Campos:2014zaa,Hernandez:2015hrt,Hernandez:2015dga,Arbelaez:2016mhg,Mantilla:2016lui, CarcamoHernandez:2016pdu,Bernal:2017xat,CarcamoHernandez:2017cwi,Mantilla:2017ijh,Abbas:2017vws,Dev:2018pjn,CarcamoHernandez:2018hst,Abbas:2018lga,CarcamoHernandez:2019pmy,CarcamoHernandez:2019vih,CarcamoHernandez:2019cbd}%
). Such models, although not necessarily excluded, may generically suffer
from large Flavor-Changing Neutral Currents (FCNCs) and from
non-observability of Higgs partners in the few-hundreds GeV mass range. In
order to explain the lepton mass hierarchy together with the highly
decoupled neutrino mass spectrum, even more additional inputs are required
on top of the SM. Due to a large number of states, such theories quickly
become cumbersome to deal with and to verify phenomenologically. Therefore,
the search for a particular model capable of explaining all the fermion mass
and mixing hierarchies in a dynamical and fully renormalizable way, while
still having it simple enough for a straightforward phenomenological
verification, becomes a challenging and demanding, but very important task
for the model-building community.

In addition, models having an extended scalar and (or) fermion sector are
motivated by the search of a theoretical explanation for the Lepton
Universality Violation (LUV) recently observed by the LHCb experiments. A
concise review of New Physics models aimed at explaining the LUV and their
possible connection to DM is provided in Ref. \cite{Vicente:2018xbv}. Some
theoretical explanations for the LUV are discussed in Refs. \cite%
{Crivellin:2015era,Crivellin:2015lwa,King:2018fcg,Bonilla:2017lsq,Barbieri:2017tuq, King:2017anf,Romao:2017qnu,Antusch:2017tud,Ko:2017quv,Ko:2017yrd,Chen:2017hir,Assad:2017iib,Angelescu:2018tyl,DiLuzio:2018zxy,Guadagnoli:2018ojc,Fornal:2018dqn,Aydemir:2018cbb,Faber:2018qon,Barman:2018jhz,Heeck:2018ntp,Grinstein:2018fgb,Falkowski:2018dsl,CarcamoHernandez:2018aon,deMedeirosVarzielas:2018bcy, Rocha-Moran:2018jzu,Hu:2018veh,Carena:2018cow,Babu:2018vrl,Allanach:2018lvl}%
.

In Ref.~\cite{CarcamoHernandez:2019cbd} we have proposed such a possible
candidate theory, capable of generating the SM fermion mass and mixing
hierarchies via a sequential loop suppression mechanism, in terms of model
parameters with no intrinsically imposed hierarchies between them. In this
framework the only fermion that acquires its mass at tree level is the heavy
top quark. Moderate and light quark masses are generated essentially at one-
or two-loop level, respectively, while light active neutrinos become massive
only via three-loop radiative seesaw mechanisms 
triggered after the electroweak symmetry breaking. We have found specific
conditions on the minimal symmetry and particle content for a theory where
this mechanism can be realized without adding the non-renormalizable
(higher-dimensional) Yukawa operators or soft family-breaking mass terms.
While such a construction is supposedly not unique, its minimality is
manifest as every field plays a relevant role for producing the observed
patterns in quark, lepton and neutrino sectors of the SM, with a required
degree of suppression between the corresponding SM parameters.


\section{Review of the extended IDM model}

\label{Sec:Extended-IDM} 

With the aim of generating the hierarchy of SM charged fermion masses via
the sequential loop suppression mechanism, proposed for the first time in
Ref.~\cite{CarcamoHernandez:2016pdu}, we consider an extension of the inert
two-Higgs doublet model (ITHDM), where the SM gauge symmetry is supplemented
by an exactly preserved $Z_{2}^{(2)}$ and spontaneously broken $Z_{2}^{(1)}$
discrete groups, and by an $U_{1X}$ gauge symmetry. The scalar sector of the
ITHDM is extended to include seven electrically neutral fields, i.e., $%
\sigma _{j}$ ($j=1,2,3$), $\rho _{k}$ ($k=1,2,3$), $\eta $ and five
electrically charged $\varphi _{k}^{+}$ ($k=1,2,3,4,5$) $SU_{2L}$ scalar
singlets. The fermion sector of the SM includes additionally six SM
gauge-singlet charged leptons $E_{jL}$ and $E_{jR}$ ($j=1,2,3$), four right
handed neutrinos $\nu _{jR}$ \ ($j=1,2,3$), $\Omega _{R}$ and twelve $SU_{2L}
$ singlet heavy quarks $T_{L}$, $T_{R}$, $\widetilde{T}_{L}$,$\ \widetilde{T}%
_{R}$, $B_{kL}$, $B_{kR}$ ($k=1,2,3,4$). It is assumed that the heavy exotic 
$T$, $\widetilde{T}$ and $B_{k}$ quarks have electric charges equal to $%
\frac{2}{3}$ and $-\frac{1}{3}$, respectively. The scalar, quark and lepton
assignments under the $SU_{3c}\times SU_{2L}\times U_{1Y}\times U_{1X}\times
Z_{2}^{(1)}\times Z_{2}^{(2)}$ symmetry are shown in Tables \ref{tab:scalars}%
, \ref{tab:quarks} and \ref{tab:leptons}, respectively. It was shown in Ref.~\cite{CarcamoHernandez:2016pdu} that with this field content and the corresponding assignments 
the gauge anomaly cancellation conditions are satisfied in our
model. 
\begin{table}[th]
\centering%
\begin{tabular}{|c||c|c|c|c|c|c|c|c|c|c|c|c|c|c|}
\hline\hline
Field & $\phi _{1}$ & $\phi _{2}$ & $\sigma _{1}$ & $\sigma _{2}$ & $\sigma
_{3}$ & $\rho _{1}$ & $\rho _{2}$ & $\rho _{3}$ & $\eta $ & $\varphi
_{1}^{+} $ & $\varphi _{2}^{+}$ & $\varphi _{3}^{+}$ & $\varphi _{4}^{+}$ & $%
\varphi _{5}^{+}$ \\ \hline
$SU_{3c}$ & $\mathbf{1}$ & $\mathbf{1}$ & $\mathbf{1}$ & $\mathbf{1}$ & $%
\mathbf{1}$ & $\mathbf{1}$ & $\mathbf{1}$ & $\mathbf{1}$ & $\mathbf{1}$ & $%
\mathbf{1}$ & $\mathbf{1}$ & $\mathbf{1}$ & $\mathbf{1}$ & $\mathbf{1}$ \\ 
\hline
$SU_{2L}$ & $\mathbf{2}$ & $\mathbf{2}$ & $\mathbf{1}$ & $\mathbf{1}$ & $%
\mathbf{1}$ & $\mathbf{1}$ & $\mathbf{1}$ & $\mathbf{1}$ & $\mathbf{1}$ & $%
\mathbf{1}$ & $\mathbf{1}$ & $\mathbf{1}$ & $\mathbf{1}$ & $\mathbf{1}$ \\ 
\hline
$U_{1Y}$ & $\frac{1}{2}$ & $\frac{1}{2}$ & $0$ & $0$ & $0$ & $0$ & $0$ & $0$
& $0$ & $1$ & $1$ & $1$ & $1$ & $1$ \\ \hline
$U_{1X}$ & $1$ & $2$ & $-1$ & $-1$ & $-2$ & $0$ & $0$ & $0$ & $1$ & $5$ & $2$
& $3$ & $2$ & $3$ \\ \hline
$Z_{2}^{(1)}$ & $1$ & $1$ & $1$ & $1$ & $-1$ & $1$ & $-1$ & $-1$ & $-1$ & $%
-1 $ & $1$ & $1$ & $-1$ & $-1$ \\ \hline
$Z_{2}^{(2)}$ & $1$ & $-1$ & $1$ & $-1$ & $-1$ & $-1$ & $-1$ & $1$ & $-1$ & $%
1$ & $1$ & $-1$ & $1$ & $1$ \\ \hline
\end{tabular}%
\caption{Scalars assignments under the $SU_{3c}\times SU_{2L}\times
U_{1Y}\times U_{1X}\times Z_{2}^{(1)}\times Z_{2}^{(2)}$ symmetry.}
\label{tab:scalars}
\end{table}
\begin{table}[th]
\centering%
\begin{tabular}{|c||c|c|c|c|c|c|c|c|c|c|c|c|c|c|c|c|c|c|c|c|c|}
\hline\hline
Field & $q_{1L}$ & $q_{2L}$ & $q_{3L}$ & $u_{1R}$ & $u_{2R}$ & $u_{3R}$ & $%
d_{1R}$ & $d_{2R}$ & $d_{3R}$ & $T_{L}$ & $T_{R}$ & $\widetilde{T}_{L}$ & $%
\widetilde{T}_{R}$ & $B_{1L}$ & $B_{1R}$ & $B_{2L}$ & $B_{2R}$ & $B_{3L}$ & $%
B_{3R}$ & $B_{4L}$ & $B_{4R}$ \\ \hline
$SU_{3c}$ & $\mathbf{3}$ & $\mathbf{3}$ & $\mathbf{3}$ & $\mathbf{3}$ & $%
\mathbf{3}$ & $\mathbf{3}$ & $\mathbf{3}$ & $\mathbf{3}$ & $\mathbf{3}$ & $%
\mathbf{3}$ & $\mathbf{3}$ & $\mathbf{3}$ & $\mathbf{3}$ & $\mathbf{3}$ & $%
\mathbf{3}$ & $\mathbf{3}$ & $\mathbf{3}$ & $\mathbf{3}$ & $\mathbf{3}$ & $%
\mathbf{3}$ & $\mathbf{3}$ \\ \hline
$SU_{2L}$ & $\mathbf{2}$ & $\mathbf{2}$ & $\mathbf{2}$ & $\mathbf{1}$ & $%
\mathbf{1}$ & $\mathbf{1}$ & $\mathbf{1}$ & $\mathbf{1}$ & $\mathbf{1}$ & $%
\mathbf{1}$ & $\mathbf{1}$ & $\mathbf{1}$ & $\mathbf{1}$ & $\mathbf{1}$ & $%
\mathbf{1}$ & $\mathbf{1}$ & $\mathbf{1}$ & $\mathbf{1}$ & $\mathbf{1}$ & $%
\mathbf{1}$ & $\mathbf{1}$ \\ \hline
$U_{1Y}$ & $\frac{1}{6}$ & $\frac{1}{6}$ & $\frac{1}{6}$ & $\frac{2}{3}$ & $%
\frac{2}{3}$ & $\frac{2}{3}$ & $-\frac{1}{3}$ & $-\frac{1}{3}$ & $-\frac{1}{3%
}$ & $\frac{2}{3}$ & $\frac{2}{3}$ & $\frac{2}{3}$ & $\frac{2}{3}$ & $-\frac{%
1}{3}$ & $-\frac{1}{3}$ & $-\frac{1}{3}$ & $-\frac{1}{3}$ & $-\frac{1}{3}$ & 
$-\frac{1}{3}$ & $-\frac{1}{3}$ & $-\frac{1}{3}$ \\ \hline
$U_{1X}$ & $0$ & $0$ & $1$ & $2$ & $2$ & $2$ & $-1$ & $-1$ & $-1$ & $1$ & $2$
& $1$ & $1$ & $0$ & $-1$ & $0$ & $-1$ & $-2$ & $-2$ & $-3$ & $-3$ \\ \hline
$Z_{2}^{(1)}$ & $1$ & $1$ & $1$ & $-1$ & $-1$ & $1$ & $-1$ & $-1$ & $-1$ & $%
1 $ & $1$ & $-1$ & $-1$ & $1$ & $1$ & $1$ & $1$ & $1$ & $1$ & $1$ & $1$ \\ 
\hline
$Z_{2}^{(2)}$ & $-1$ & $-1$ & $-1$ & $-1$ & $-1$ & $-1$ & $-1$ & $-1$ & $-1$
& $1$ & $1$ & $-1$ & $-1$ & $1$ & $1$ & $1$ & $1$ & $1$ & $1$ & $-1$ & $-1$
\\ \hline
\end{tabular}%
\caption{Quark assignments under the $SU_{3c}\times SU_{2L}\times
U_{1Y}\times U_{1X}\times Z_{2}^{(1)}\times Z_{2}^{(2)}$ symmetry.}
\label{tab:quarks}
\end{table}
\begin{table}[th]
\centering%
\begin{tabular}{|c||c|c|c|c|c|c|c|c|c|c|c|c|c|c|c|c|c|c|}
\hline\hline
Field & $l_{1L}$ & $l_{2L}$ & $l_{3L}$ & $l_{1R}$ & $l_{2R}$ & $l_{3R}$ & $%
E_{1L}$ & $E_{1R}$ & $E_{2L}$ & $E_{2R}$ & $E_{3L}$ & $E_{3R}$ & $\nu _{1R}$
& $\nu _{2R}$ & $\nu _{3R}$ & $\Omega _{1R}$ & $\Omega _{2R}$ & $\Psi _{R}$
\\ \hline
$SU_{3c}$ & $\mathbf{1}$ & $\mathbf{1}$ & $\mathbf{1}$ & $\mathbf{1}$ & $%
\mathbf{1}$ & $\mathbf{1}$ & $\mathbf{1}$ & $\mathbf{1}$ & $\mathbf{1}$ & $%
\mathbf{1}$ & $\mathbf{1}$ & $\mathbf{1}$ & $\mathbf{1}$ & $\mathbf{1}$ & $%
\mathbf{1}$ & $\mathbf{1}$ & $\mathbf{1}$ & $\mathbf{1}$ \\ \hline
$SU_{2L}$ & $\mathbf{2}$ & $\mathbf{2}$ & $\mathbf{2}$ & $\mathbf{1}$ & $%
\mathbf{1}$ & $\mathbf{1}$ & $\mathbf{1}$ & $\mathbf{1}$ & $\mathbf{1}$ & $%
\mathbf{1}$ & $\mathbf{1}$ & $\mathbf{1}$ & $\mathbf{1}$ & $\mathbf{1}$ & $%
\mathbf{1}$ & $\mathbf{1}$ & $\mathbf{1}$ & $\mathbf{1}$ \\ \hline
$U_{1Y}$ & $-\frac{1}{2}$ & $-\frac{1}{2}$ & $-\frac{1}{2}$ & $-1$ & $-1$ & $%
-1$ & $-1$ & $-1$ & $-1$ & $-1$ & $-1$ & $-1$ & $0$ & $0$ & $0$ & $0$ & $0$
& $0$ \\ \hline
$U_{1X}$ & $0$ & $-3$ & $0$ & $-3$ & $-6$ & $-3$ & $-3$ & $-2$ & $-6$ & $-5$
& $-3$ & $-2$ & $2$ & $-1$ & $2$ & $-1$ & $1$ & $0$ \\ \hline
$Z_{2}^{(1)}$ & $1$ & $-1$ & $1$ & $1$ & $-1$ & $1$ & $-1$ & $-1$ & $-1$ & $%
-1$ & $1$ & $1$ & $1$ & $-1$ & $1$ & $-1$ & $-1$ & $1$ \\ \hline
$Z_{2}^{(2)}$ & $-1$ & $-1$ & $-1$ & $-1$ & $-1$ & $-1$ & $1$ & $1$ & $1$ & $%
1$ & $1$ & $1$ & $1$ & $1$ & $1$ & $-1$ & $1$ & $1$ \\ \hline
\end{tabular}%
\caption{Lepton charge assignments under the $SU_{3c}\times SU_{2L}\times
U_{1Y}\times U_{1X}\times Z_{2}^{(1)}\times Z_{2}^{(2)}$ symmetry.}
\label{tab:leptons}
\end{table}
Let us note that the SM Higgs doublet, i.e., $\phi _{1}$, as well as the SM
scalar singlets $\sigma _{1}$ and $\rho _{3}$ are the only scalar fields
neutral under the preserved $Z_{2}^{(2)}$ discrete symmetry. Since the $%
Z_{2}^{(2)}$ symmetry remains unbroken, the SM Higgs doublet $\phi _{1}$ and
the SM scalar singlets $\sigma _{1}$ and $\rho _{3}$ are the only scalar
fields which acquire nonvanishing vacuum expectation values. The SM scalar
singlet $\sigma _{1}$ is required to spontaneously break the $U_{1X}$ local
symmetry, whereas the scalar singlet $\rho _{3}$ spontaneously breaks the $%
Z_{2}^{(1)}$ discrete symmetry, due to its nontrivial $Z_{2}^{(1)}$ charge. 

In the following we provide a brief justification for introducing different
particles in our model. It is worth mentioning that the set of $SU_{2L}$%
-singlet heavy quarks $T_{L}$, $T_{R}$, $B_{iL}$, $B_{iR}$ ($i=1,2,3$)
represents the minimal amount of exotic quark degrees of freedom needed to
implement the one-loop radiative seesaw mechanism that gives rise to the
charm, bottom and strange quark masses. In addition, to implement this
one-loop radiative seesaw mechanism, one needs the extra $\phi _{2}$ scalar
doublet, the gauge singlet scalars $\rho _{2}$, $\rho _{3}$\ and $\eta $, 
charged under the preserved $Z_{2}^{(2)}$ symmetry as well as the scalar
singlets $\sigma _{1}$ and $\rho _{3}$ that spontaneously break the $U_{1X}$
and $Z_{2}^{(1)}$ symmetries, respectively. Furthermore, in order to ensure
the radiative seesaw mechanism responsible for the generation of the up and
down quark masses at two-loop level, the $SU_{2L}$ singlet heavy quarks $%
\widetilde{T}_{L}$, $\widetilde{T}_{R}$, $B_{4L}$, $B_{4R}$, as well as the
electrically neutral, $\sigma _{3}$, $\rho _{2}$, and electrically charged, $%
\varphi _{1}^{+}$, $\varphi _{2}^{+}$ scalar $SU_{2L}$-singlets should also
be present in the particle spectrum. Furthermore, the generation of one-loop
tau and muon masses is mediated by the electrically charged weak-singlet
leptons $E_{rL}$ and $E_{rR}$ ($r=2,3$), by the inert scalar $SU_{2L}$%
-doublet, $\phi _{2}$, and by the $SU_{2L}$-singlets $\sigma _{2}$, $\rho
_{1}$. On the other hand, to induce a non-zero electron mass at two-loop
level, and extra weak-singlet charged $E_{1}$ and neutral $\nu _{mR}$ ($m=1,3
$), \ leptons $\Psi _{R}$ as well as the electrically charged scalar
singlets $\varphi _{1}^{\pm }$, $\varphi _{k}^{\pm }$\ \ ($k=3,4,5$) would
be required for this purpose. Moreover, the three-loop radiative seesaw
mechanism responsible for the generation of the light active neutrino masses
is mediated by the right-handed neutrinos $\nu _{jR}$ ($j=1,2,3$), $\Omega
_{R}$, as well as by the inert scalar $SU_{2L}$ doublet $\phi _{2}$ and the $%
SU_{2L}$-singlet $\sigma _{2}$. More details for the choice of the
aforementioned particle content and symmetries are provided in our previous
work in Ref. \cite{CarcamoHernandez:2019cbd}. %

\begin{figure}[h]
\resizebox{18cm}{22cm} {\includegraphics{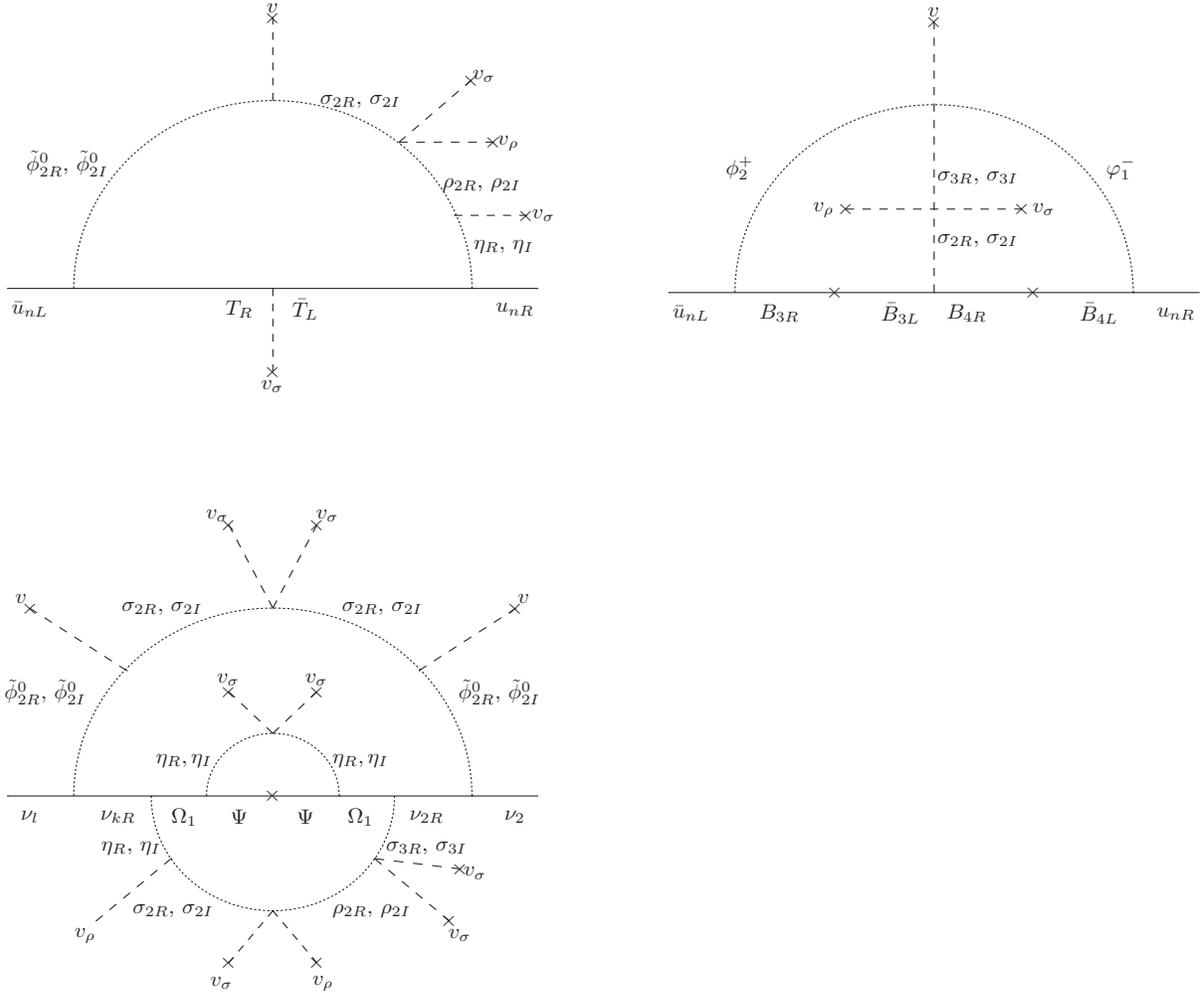}} \vspace{-7.5cm%
}
\caption{Some of the one-, two- and three-loop Feynman diagrams contributing
to the entries of the SM fermion mass matrices. Here, $n,m=1,2$, $l,n=1,3$.}
\label{Loopdiagrams}
\end{figure}

With the above specified particle content, the following Yukawa interactions
and exotic fermion mass terms are present at renormalizable level, invariant
under the $SU_{3c}\times SU_{2L}\times U_{1Y}\times U_{1X}\times
Z_{2}^{(1)}\times Z_{2}^{(2)}$ symmetry:
\begin{eqnarray}
\tciLaplace _{\mathrm{F}} &=&y_{3j}^{\left( u\right) }\overline{q}_{3L}%
\widetilde{\phi }_{1}u_{3R}+\sum_{n=1}^{2}x_{n}^{\left( u\right) }\overline{q%
}_{nL}\widetilde{\phi }_{2}T_{R}+\sum_{n=1}^{2}z_{j}^{\left( u\right) }%
\overline{T}_{L}\eta ^{\ast }u_{nR}+y_{T}\overline{T}_{L}\sigma _{1}T_{R}+m_{%
\widetilde{T}}\overline{\widetilde{T}}_{L}\widetilde{T}_{R}+x^{\left(
T\right) }\overline{T}_{L}\rho _{2}\widetilde{T}_{R}  \notag \\
&&+\sum_{n=1}^{2}x_{n}^{\left( d\right) }\overline{q}_{3L}\phi
_{2}B_{nR}+\sum_{n=1}^{2}\sum_{j=1}^{3}y_{nj}^{\left( d\right) }\overline{B}%
_{nL}\eta d_{jR}+\sum_{j=1}^{3}z_{j}^{\left( d\right) }\overline{B}_{3L}\eta
^{\ast }d_{jR}+\sum_{n=1}^{2}w_{n}^{\left( u\right) }\overline{B}%
_{4L}\varphi _{1}^{-}u_{nR}  \notag \\
&&+\sum_{k=3}^{4}m_{B_{k}}\overline{B}_{kL}B_{kR}+\sum_{n=1}^{2}x_{n}^{%
\left( d\right) }\overline{q}_{nL}\phi
_{2}B_{3R}+\sum_{n=1}^{2}\sum_{m=1}^{2}y_{nm}^{\left( B\right) }\overline{B}%
_{nL}\sigma _{1}^{\ast }B_{mR}+z^{\left( B\right) }\overline{B}_{3L}\sigma
_{2}^{\ast }B_{4R}+\sum_{j=1}^{3}w_{j}^{\left( d\right) }\overline{%
\widetilde{T}}_{L}\varphi _{2}^{+}d_{jR}  \notag \\
&&+\sum_{k=1,3}x_{k3}^{\left( l\right) }\overline{l}_{kL}\phi
_{2}E_{3R}+\sum_{k=1,3}y_{3k}^{\left( l\right) }\overline{E}_{3L}\rho
_{1}l_{kR}+x_{22}^{\left( l\right) }\overline{l}_{2L}\phi
_{2}E_{2R}+y_{22}^{\left( l\right) }\overline{E}_{2L}\rho _{1}l_{2R}  \notag
\\
&&+\sum_{i=1}^{3}y_{i}^{\left( E\right) }\overline{E}_{iL}\sigma _{1}^{\ast
}E_{iR}+x_{2}^{\left( \nu \right) }\overline{l}_{2L}\widetilde{\phi }_{2}\nu
_{2R}+\sum_{k=1,3}z_{k}^{\left( l\right) }\overline{\Psi _{R}^{C}}\varphi
_{3}^{+}l_{kR}+\sum_{k=1,3}z_{k}^{\left( \nu \right) }\overline{E}%
_{1L}\varphi _{1}^{-}\nu _{kR}+z^{\left( E\right) }\overline{\Psi _{R}^{C}}%
\varphi _{4}^{+}E_{1R}  \notag \\
&&+\sum_{k=1,3}\sum_{n=1,3}x_{kn}^{\left( \nu \right) }\overline{l}_{kL}%
\widetilde{\phi }_{2}\nu _{nR}+\sum_{k=1,3}y_{k}^{\left( \Omega \right) }%
\overline{\Omega _{1R}^{C}}\eta ^{\ast }\nu _{kR}+y^{\left( \Omega \right) }%
\overline{\Omega _{1R}^{C}}\sigma _{3}^{\ast }\nu _{2R}  \notag \\
\label{eq:Yukawas}
&&+x_{1}^{\left( \Psi \right) }\overline{\Omega _{1R}^{C}}\eta \Psi
_{R}+x_{2}^{\left( \Psi \right) }\overline{\Omega _{2R}^{C}}\eta ^{\ast
}\Psi _{R}+z_{\Omega }\overline{\Omega _{1R}^{C}}\sigma _{2}^{\ast }\Omega
_{2R}+m_{\Psi }\overline{\Psi _{R}^{C}}\Psi _{R}+h.c.\,,
\end{eqnarray}
where the dimensionless couplings are $\mathcal{O}(1)$ parameters. From the
quark Yukawa terms it follows that the top quark mass only arises from the
interaction with the SM Higgs doublet $\phi _{1}$. After the spontaneous
breaking of the SM electroweak symmetry, the observed hierarchy of SM
fermion masses arises by a sequential loop suppression, such that we have:
tree-level top quark mass; one-loop bottom, strange, charm, tau and muon
masses; two-loop masses for the up, down quarks as well as for the electron.
Furthermore, light active neutrinos get their masses from a three-loop level
radiative seesaw mechanism. Some of the one-, two- and three-loop Feynman
diagrams contributing to the entries of the SM fermion mass matrices are
shown in Figure \ref{Loopdiagrams}. More details are given in our previous
work. 


\section{Constraints on the $Z^{\prime }$ mass, couplings and production at
the LHC}


In this section, we discuss the constraints on the $Z^{\prime }$ mass and
couplings in our model that emerge due to the $2.6\sigma $ lepton
universality anomaly expressed as the ratio $R_{K}=\frac{Br\left(
B\rightarrow K\mu ^{+}\mu^{-}\right) }{Br\left( B\rightarrow
Ke^{+}e^{-}\right) }$ measured by the LHCb collaboration. In addition, we
will determine the LEP constraint on the $M_{Z^{\prime }}/g_{X}$ ratio. As
we will show below, in our model the lepton universality violation is a
consequence of the non-universal $U_{1X}$ charge assignments of the
fermionic fields. From the $U_{1X}$ assignments for fermions, we find the
following $Z^{\prime }$ interactions with the SM fermions: 
\begin{eqnarray}
\tciLaplace _{Z`} &=&g_{X}\overline{q}_{3L}\gamma ^{\mu
}q_{3L}Z_{\mu}^{\prime }+2g_{X}\sum_{j=1}^{3}\overline{u}_{jR}\gamma ^{\mu
}u_{jR}Z_{\mu}^{\prime }- g_{X}\sum_{j=1}^{3}\overline{d}_{jR}\gamma ^{\mu
}d_{jR}Z_{\mu}^{\prime }  \notag \\
&&-3g_{X}\overline{l}_{2L}\gamma ^{\mu }l_{2L}Z_{\mu }^{\prime }-6g_{X}%
\overline{l}_{2R}\gamma ^{\mu }l_{2R}Z_{\mu }^{\prime }-3g_{X}\sum_{k=1,3} 
\overline{l}_{kR}\gamma ^{\mu }l_{kR}Z_{\mu }^{\prime } \,.
\end{eqnarray}
Then the non-universal $Z^{\prime }$ interactions with the SM fermions given
above lead to the following effective Hamiltonian, where the fermionic
fields are given in the physical basis: 
\begin{eqnarray}
\Delta H_{eff} &=&-\frac{g_{X}^{2}}{M_{Z^{\prime }}^{2}}\left( V_{DL}^{\ast
}\right) _{32}\left( V_{DL}\right) _{33}x_{q_{3L}}\sum_{j=1}^{3}\left[
x_{l_{jL}}\left( \overline{s}\gamma ^{\mu }P_{L}b\right) \left( \overline{l}%
_{jL}\gamma ^{\mu }l_{jL}\right) +x_{l_{jR}}\left( \overline{s}\gamma ^{\mu}
P_{L}b\right) \left( \overline{l}_{jR}\gamma ^{\mu }l_{jR}\right) \right] 
\notag \\
&=&-\frac{g_{X}^{2}}{M_{Z^{\prime }}^{2}}\left( V_{DL}^{\ast
}\right)_{32}\left( V_{DL}\right) _{33}\left[ -3\left( \overline{s}\gamma
^{\mu} P_{L}b\right) \left( \overline{l}_{2L}\gamma ^{\mu }l_{2L}\right)
-6\left( \overline{s}\gamma ^{\mu }P_{L}b\right) \left( \overline{l}%
_{2R}\gamma^{\mu} l_{2R}\right) \right.  \notag \\
&&-\left. 3\left( \overline{s}\gamma ^{\mu }P_{L}b\right) \left( \overline{l}%
_{1R}\gamma ^{\mu }l_{1R}\right) -3\left( \overline{s}\gamma ^{\mu}
P_{L}b\right) \left( \overline{l}_{3R}\gamma ^{\mu }l_{3R}\right) \right] 
\notag \\
&\supset &\frac{9g_{X}^{2}}{2M_{Z^{\prime }}^{2}}\left(
V_{DL}^{\ast}\right)_{32}\left( V_{DL}\right) _{33}\left( \overline{s}\gamma
^{\mu} P_{L}b\right) \left( \overline{\mu }\gamma ^{\mu }\mu \right) \,,
\end{eqnarray}
where the following relations have been taken into account: 
\begin{eqnarray}
\widetilde{M}_{f} &=&\left( M_{f}\right) _{diag}=V_{fL}^{\dagger}M_{f}V_{fR},%
\hspace{1cm}\hspace{1cm}f_{\left( L,R\right) }=V_{f\left(L,R\right) } 
\widetilde{f}_{\left( L,R\right) },  \notag \\
\overline{f}_{iL}\left( M_{f}\right) _{ij}f_{jR} &=&\overline{\widetilde{f}}%
_{kL}\left( V_{fL}^{\dagger }\right) _{ki}\left( M_{f}\right)_{ij}
\left(V_{fR}\right) _{jl}\widetilde{f}_{lR}=\overline{\widetilde{f}}%
_{kL}\left(V_{fL}^{\dagger }M_{f}V_{fR}\right) _{kl}\widetilde{f}_{lR}= 
\overline{\widetilde{f}}_{kL}\left( \widetilde{M}_{f}\right) _{kl}\widetilde{%
f}_{lR}=m_{fk}\overline{\widetilde{f}}_{kL}\widetilde{f}_{kR},  \notag \\
k &=&1,2,3 \,.
\end{eqnarray}
Here, $\widetilde{f}_{k\left( L,R\right) }$ and $f_{k\left( L,R\right) }$ ($%
k=1,2,3$) are the SM fermionic fields in the mass and interaction bases,
respectively. 
\begin{table}[tbp]
\centering
\begin{tabular}{|c|c|}
\hline
Parameter & $\frac{\Delta C_{9}^{\mu \mu }}{C_{9}^{SM}}$ \\ \hline
Best fit & $-0.21$ \\ \hline
$1\sigma$ range & $-0.27$ up to $-0.13$ \\ \hline
$2\sigma$ range & $-0.32$ up to $-0.08$ \\ \hline
\end{tabular}%
\caption{Constraints on the $C_{9}^{\protect\mu \protect\mu }$ Wilson
coefficient from the LHCb data. Taken from Ref.~\protect\cite{Hurth:2016fbr}%
. }
\label{C9}
\end{table}

Let us note that the $R_{K}$ anomaly results from a shift in the Wilson
coefficient $C_{9}^{\mu \mu }$ appearing in the following $\Delta B=1$
effective Hamiltonian: 
\begin{equation}
\Delta H_{eff}=-\frac{G_{F}\alpha _{em}V_{tb}V_{ts}^{\ast }}{\sqrt{2}\pi }
\sum_{\widetilde{l}=e,\mu ,\tau }C_{9}^{\widetilde{l}\widetilde{l}}\left( 
\overline{s}\gamma ^{\mu }P_{L}b\right) \left( \overline{\widetilde{l}}
\gamma ^{\mu }\widetilde{l}\right) \,.
\end{equation}
Then, our model predicts the following correction to the $C_{9}^{\mu \mu }$
coefficient relative to its SM value: 
\begin{equation}
\Delta C_{9}^{\mu \mu }=-\frac{9g_{X}^{2}}{2M_{Z^{\prime }}^{2}}\left(
V_{DL}^{\ast }\right) _{32}\left( V_{DL}\right) _{33}\frac{\sqrt{2}\pi }{
G_{F}\alpha _{em}V_{tb}V_{ts}^{\ast }}\simeq -\frac{9g_{X}^{2}}{
2M_{Z^{\prime }}^{2}}\frac{\sqrt{2}\pi }{G_{F}\alpha_{em}} \,.
\end{equation}

On the other hand, the LHCb data provide the constraints on the $%
C_{9}^{\mu\mu }$ coefficient given in Table \ref{C9}. Requiring for the
correction to the $C_{9}^{\mu \mu }$ coefficient predicted by our model to
be inside the $1\sigma $ and $2\sigma $ experimentally allowed ranges, we
find the constraints for the $M_{Z^{\prime }}/g_{X}$ ratio: 
\begin{equation}
14\;\mbox{TeV}<\frac{M_{Z^{\prime }}}{g_{X}}<20\;\mbox{TeV}\;\; \mbox{at 1}%
\sigma, \hspace{1cm}\hspace{1cm}13\;\mbox{TeV}<\frac{M_{Z^{\prime }}}{g_{X}}%
<26\; \mbox{TeV}\;\; \mbox{at 2}\sigma \,.
\end{equation}

With respect to the LEP bounds on the $M_{Z^{\prime }}/g_{X}$ ratio, it is
worth mentioning that the tightest constraint arises from the $%
e^{+}e^{-}\rightarrow \mu ^{+}\mu ^{-}$ measurement at LEP. Using the
effective leptonic interactions 
\begin{eqnarray}
\tciLaplace _{eff} &=&-\frac{g_{X}^{2}}{M_{Z^{\prime }}^{2}}\sum_{j=1}^{3}%
\left[ x_{l_{1L}}x_{l_{jL}}\left( \overline{l}_{1}\gamma ^{\mu
}P_{L}l_{1}\right) \left( \overline{l}_{jL}\gamma ^{\mu }l_{jL}\right)
+x_{l_{1L}}x_{l_{jR}}\left( \overline{l}_{1}\gamma ^{\mu }P_{L}l_{1}\right)
\left( \overline{l}_{jR}\gamma ^{\mu }l_{jR}\right) \right]  \notag \\
&&-\frac{g_{X}^{2}}{M_{Z^{\prime }}^{2}}\sum_{j=1}^{3}\left[
x_{l_{1R}}x_{l_{jL}}\left( \overline{l}_{1}\gamma ^{\mu }P_{R}l_{1}\right)
\left( \overline{l}_{jL}\gamma ^{\mu }l_{jL}\right)
+x_{l_{1R}}x_{l_{jR}}\left( \overline{l}_{1}\gamma ^{\mu }P_{R}l_{1}\right)
\left( \overline{l}_{jR}\gamma ^{\mu }l_{jR}\right) \right] \,,
\end{eqnarray}
we find that the $e^{+}e^{-}\rightarrow \mu ^{+}\mu ^{-}$ measurement at LEP
imposes the following limit \cite{Schael:2013ita}: 
\begin{equation}
\frac{2M_{Z^{\prime }}}{g_{X}\sqrt{%
x_{l_{1L}}x_{l_{2L}}+x_{l_{1R}}x_{l_{2R}}+x_{l_{1R}}x_{l_{2L}}+x_{l_{1L}}x_{l_{2R}}%
}} > 4.6\;\mbox{TeV} \,,
\end{equation}
which for the leptonic charge assignments of our model takes the form: 
\begin{equation}
\frac{2M_{Z^{\prime }}}{3\sqrt{3}g_{X}}\simeq 0.38\frac{M_{Z^{\prime }}}{%
g_{X}}>4.6\;\mbox{TeV} \,.
\end{equation}
The latter yields the following lower bound on the $M_{Z^{\prime }}/g_{X}$
ratio: 
\begin{equation}
\frac{M_{Z^{\prime }}}{g_{X}}>12 \;\mbox{TeV} \,.
\end{equation}
In what follows, we proceed with computing the total cross section for
production of a heavy $Z^{\prime }$ gauge boson at the LHC via a {\
Drell-Yan (DY) mechanism.} In this computation, we consider the dominant
contribution due to the parton distribution functions of the light up, down
and strange quarks, so that the total $Z^{\prime }$ production cross section
via quark-antiquark annihilation in proton-proton collisions with
center-of-mass energy $\sqrt{S}$ reads: 
\begin{eqnarray}
\sigma_{pp\rightarrow Z^{\prime }}^{\mathrm{DY}}\left( S\right) &=&\frac{%
g^{2}\pi }{6c_{W}^{2}S}\left\{ 4g_{X}^{2}\int_{\ln \sqrt{\frac{ M_{Z^{\prime
}}^{2}}{S}}}^{-\ln \sqrt{\frac{M_{Z^{\prime }}^{2}}{S}} }f_{p/u}\left( \sqrt{%
\frac{M_{Z^{\prime }}^{2}}{S}}e^{y},\mu ^{2}\right) f_{p/\overline{u}}\left( 
\sqrt{\frac{M_{Z^{\prime }}^{2}}{S}}e^{-y},\mu^{2}\right) dy\right.  \notag
\\
&&+\left. g_{X}^{2}\int_{\ln \sqrt{\frac{M_{Z^{\prime }}^{2}}{S}}}^{-\ln 
\sqrt{\frac{M_{Z^{\prime }}^{2}}{S}}}f_{p/d}\left( \sqrt{\frac{M_{Z^{\prime
}}^{2}}{S}}e^{y},\mu ^{2}\right) f_{p/\overline{d}}\left( \sqrt{\frac{%
M_{Z^{\prime }}^{2}}{S}}e^{-y},\mu ^{2}\right) dy\right.  \notag \\
&&+\left. g_{X}^{2}\int_{\ln \sqrt{\frac{M_{Z^{\prime }}^{2}}{S}}}^{-\ln 
\sqrt{\frac{M_{Z^{\prime }}^{2}}{S}}}f_{p/s}\left( \sqrt{\frac{M_{Z^{\prime
}}^{2}}{S}}e^{y},\mu ^{2}\right) f_{p/\overline{s}}\left( \sqrt{\frac{%
M_{Z^{\prime }}^{2}}{S}}e^{-y},\mu ^{2}\right) dy\right\} \,,
\end{eqnarray}
where $f_{p/u}\left( x_1,\mu ^2 \right)$ ($f_{p/\overline{u}}\left(x_2,\mu
^2 \right)$), $f_{p/d}\left(x_1,\mu ^2 \right)$ ($f_{p/\overline{d}}\left(
x_2,\mu ^2 \right)$) and $f_{p/s}\left(x_1,\mu ^2 \right)$ ($f_{p/\overline{s%
}}\left( x_2,\mu ^2 \right)$) are the distributions of the light up, down
and strange quarks (antiquarks) in the proton, respectively, which carry
momentum fractions $x_1$ ($x_2$) of the proton. Here, $\mu=m_{Z^\prime }$ is
the corresponding factorization scale. 
\begin{figure}[tbh]
\resizebox{12cm}{8cm}{\includegraphics{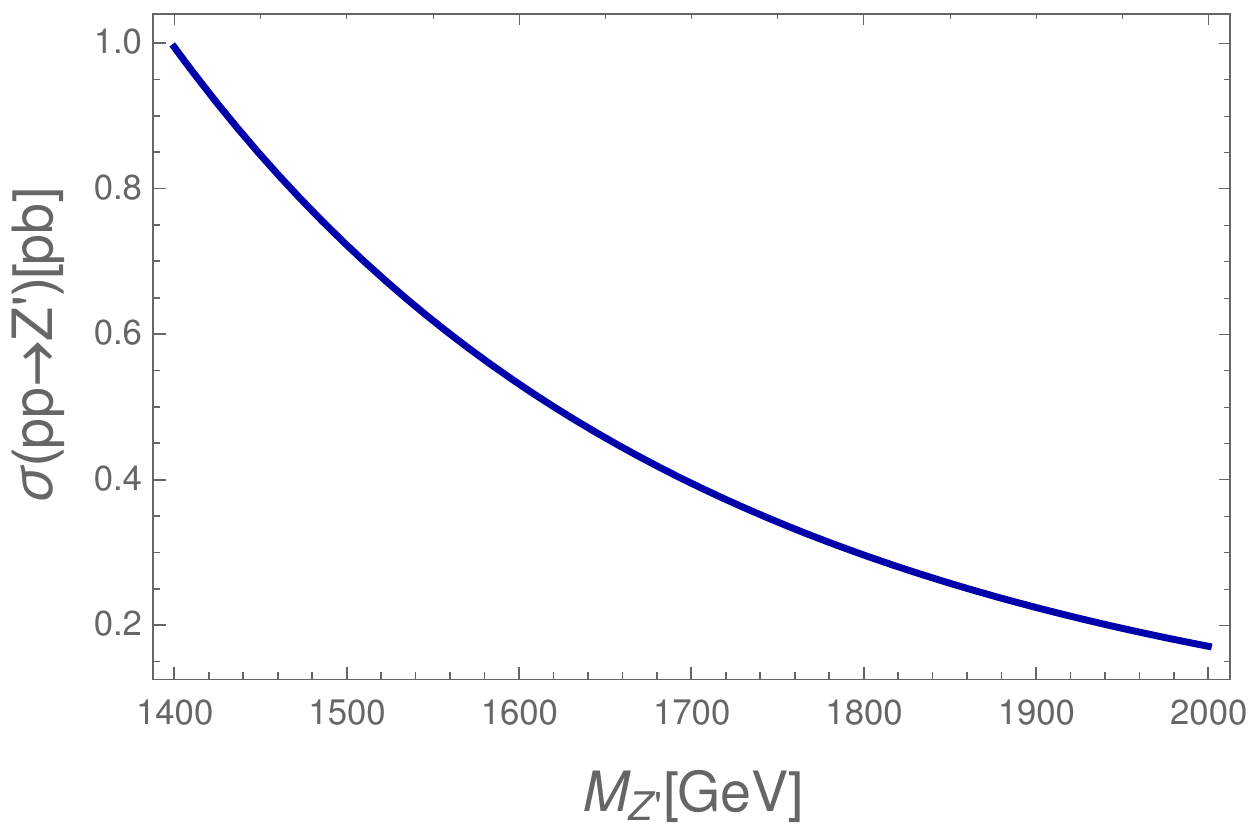}}\vspace{0cm}
\caption{The total $Z^{\prime }$ production cross section via the DY
mechanism at the LHC for $\protect\sqrt{S}=13$ TeV and $g_{X}=0.1$ as a
function of the $Z^{\prime }$ mass.}
\label{qqtoZprime}
\end{figure}

\begin{figure}[tbh]
\resizebox{12cm}{8cm}{\includegraphics{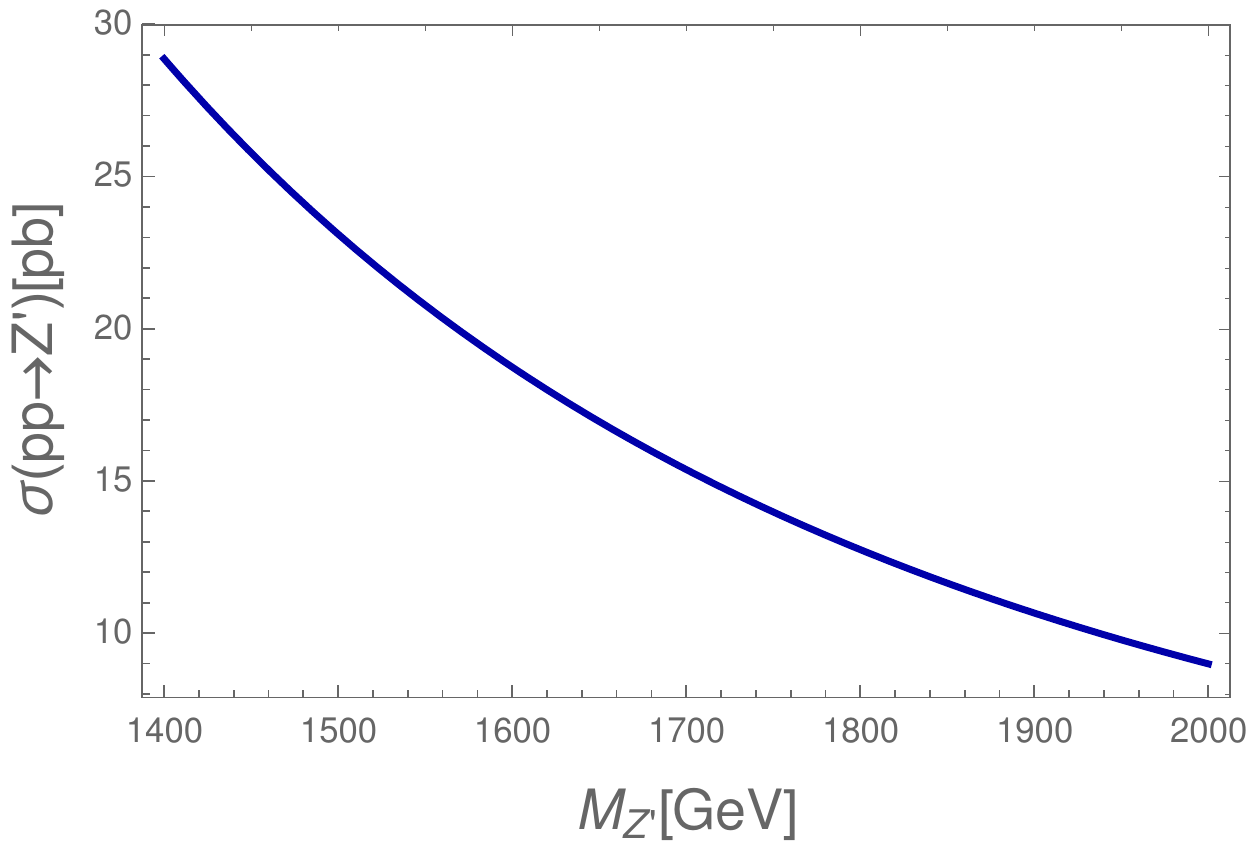}}\vspace{0cm}
\caption{The total $Z^{\prime }$ production cross section via the DY
mechanism at a future $pp$ collider for $\protect\sqrt{S}=100$ TeV and $%
g_{X}=0.1$ as a function of the $Z^{\prime }$ mass.}
\label{qqtoZprimefor100TeV}
\end{figure}

Figure \ref{qqtoZprime} displays the total $Z^{\prime }$ production cross
section via the DY mechanism at the LHC for $\sqrt{S}=13$ TeV and $g_{X}=0.1$
as a function of the $Z^{\prime }$ mass. The latter is varied from $1.4$ TeV
up to $2$ TeV to satisfy the LEP constraint as well as the constraints
imposed by the $2.6\sigma $ anomaly in lepton universality. For such as a
region of $Z^{\prime }$ masses, we find that the total production cross
section is found to be $0.2-1$ pb. On the other hand, at a future $100$ TeV
proton-proton collider this cross section gets significantly enhanced
reaching values of $9-29$ pb in the same mass interval, as indicated in
Figure \ref{qqtoZprimefor100TeV}. 
Note that the dominant $Z^{\prime }$ production channel in pp collisions is via the Drell-Yan process $q\bar q\rightarrow Z^{\prime }$, $q=u,d,s$ (for the production 
cross section see Eq. (12)). The produced $Z^{\prime }$ then can decay into a leptonic pair $Z^{\prime }\rightarrow l\bar l$ which is a standard search 
channel for $Z^{\prime }$ at the LHC. Non-observation of such a decay channel at LEP-II yields the lower bound 
$M_{^{\prime }}/g_X > 12$ TeV (see Eq. (11)). The hadron collider observables of $Z^{\prime }$ in hadronic and leptonic channels 
in the current model requires a dedicated analysis of the experimental bounds in each of the $Z^{\prime }$ decay channels, 
and should be left for future studies.

Finally, to close this section, we justify why our model is safe against Flavor Changing Neutral Currents (FCNCs) constraints.
Since the model contains two electroweak doublet Higgs scalars $\phi_{1,2}$ we are required to take special care of  this issue.  
%
Our model automatically implements the alignment limit for the lightest 125 GeV Higgs boson, because all other scalar states decouple in the mass spectrum and, hence, are very heavy by default. This means
the SM-like Higgs boson does not have tree-level FCNCs while such contributions from the heavier scalars
are strongly suppressed by their large mass scale. While a detailed study of the FCNC constraints goes beyond the scope of the present work, we can resort to the Glashow-Weinberg-Paschos the theorem  \cite{Glashow:1976nt,Paschos:1976ay} in order to justify nonexistence of FCNCs in our model. 
This theorem states that there will be no tree-level FCNC coming from the scalar sector, if all right-handed 
fermions of a given electric charge couple to only one of the doublets. As seen from Eq.~(\ref{eq:Yukawas}) this condition is satisfied in our model.
So, despite of an obvious mass suppression, any possible FCNC corrections would emerge at a loop level only, guarantying the model to be safe with respect to the corresponding phenomenological constraints. Finally, any possible FCNC from the $Z'$ mediation would be strongly suppressed by its large mass 
scale compared to the EW one, i.e. $m_{Z'} >12$ TeV (for $g_X=1$), according to the LEP constraint.


\section{Muon anomalous magnetic moment}


In this section, we will determine the constraints on the parameter space of
our model imposed by the experimental measurements of the muon anomalous
magnetic moment. The latter receives one-loop contributions from vertex
diagrams involving the $Z^{\prime }$ exchange as well as the exchanges of
the heavy $Z_{2}^{\left( 2\right) }$ charged neutral scalars $\func{Re}%
\left( \rho _{1}\right) $, $\func{Re}\left( \phi _{2}^{0}\right) $, $\func{Im%
}\left( \phi _{2}^{0}\right) $, $\func{Im}\left( \rho _{1}\right) $, that
couple to the charged exotic lepton $E_{2}$. The scalar contributions to the
muon anomalous magnetic moment include the Yukawa interactions $\overline{E}%
_{2L}\rho _{1}l_{2R}$ and $\overline{l}_{2L}\phi _{2}E_{2R}$ as well as the
trilinear scalar interactions such as $\rho _{2}\left( \phi _{1}\cdot \phi
_{2}^{\dagger }\right) \sigma _{1}^{\ast }$ giving rise to the $\phi _{2}^{0}
$-$\rho _{2}$ mixing, which is crucial to generate those contributions.

In view of a huge amount of free parameters in the scalar potential of our
model (which is shown explicitly in our previous work in Ref. \cite%
{CarcamoHernandez:2019cbd}), for the sake of simplicity, here we work with a
simplified benchmark scenario where $\func{Re}\left( \rho _{1}\right) $ ($%
\func{Im}\left( \rho _{1}\right) $) and $\func{Re}\left( \phi
_{2}^{0}\right) $ ($\func{Im}\left( \phi _{2}^{0}\right) $) mix between
themselves only and do not mix with other scalar fields. In this scenario,
we have the following relations: 
\begin{equation}
\left( 
\begin{array}{c}
H_{1} \\ 
H_{2}%
\end{array}%
\right) =\left( 
\begin{array}{cc}
\cos \theta _{S} & \sin \theta _{S} \\ 
-\sin \theta _{S} & \cos \theta _{S}%
\end{array}%
\right) \left( 
\begin{array}{c}
\func{Re}\left( \rho _{1}\right) \\ 
\func{Re}\left( \phi _{2}^{0}\right)%
\end{array}%
\right) ,\hspace{1cm}\hspace{1cm}\left( 
\begin{array}{c}
A_{1} \\ 
A_{2}%
\end{array}%
\right) =\left( 
\begin{array}{cc}
\cos \theta _{P} & \sin \theta _{P} \\ 
-\sin \theta _{P} & \cos \theta _{P}%
\end{array}%
\right) \left( 
\begin{array}{c}
\func{Im}\left( \rho _{1}\right) \\ 
\func{Im}\left( \phi _{2}^{0}\right)%
\end{array}%
\right)
\end{equation}%
where $H_{1}$, $H_{2}$ are the physical CP-even scalars whereas $A_{1}$ and $%
A_{2}$ are the CP-odd scalars in the physical basis. In addition, without
any loss of generality we set $\theta _{S}=\theta _{P}=\theta $ and $%
y_{22}^{\left( l\right) }=x_{22}^{\left( l\right) }=y$. Then, the muon
anomalous magnetic moment in this scenario reads: 
\begin{eqnarray}
\Delta a_{\mu } &=&y^{2}\frac{m_{\mu }^{2}}{8\pi ^{2}}\left[ I_{S}\left(
m_{E_{2}},m_{H_{1}}\right) -I_{S}\left( m_{E_{2}},m_{H_{2}}\right)
+I_{P}\left( m_{E_{2}},m_{A_{1}}\right) -I_{P}\left(
m_{E_{2}},m_{A_{2}}\right) \right] \sin \theta \cos \theta +\frac{m_{\mu
}^{2}}{8\pi ^{2}M_{Z^{\prime }}^{2}}I_{V}\left( M_{Z^{\prime }}\right) \,, 
\notag \\
&&
\end{eqnarray}%
where the loop integrals are given by \cite{Diaz:2002uk,Kelso:2014qka}: 
\begin{eqnarray}
I_{S\left( P\right) }\left( m_{E},m\right) &=&\int_{0}^{1}\frac{x^{2}\left(
1-x\pm \frac{m_{E}}{m_{\mu }}\right) }{m_{\mu }^{2}x^{2}+\left(
m_{E}^{2}-m_{\mu }^{2}\right) x+m^{2}\left( 1-x\right) }dx,  \notag \\
I_{V}\left( M_{Z^{\prime }}\right) &=&\int_{0}^{1}\frac{g_{V}^{2}P_{V}\left(
x\right) +g_{A}^{2}P_{A}\left( x\right) }{\left( 1-x\right) \left( 1-\frac{%
m_{\mu }^{2}}{M_{Z^{\prime }}^{2}}x\right) +\frac{m_{\mu }^{2}}{M_{Z^{\prime
}}^{2}}x}dx,\hspace{1cm}  \notag \\
P_{V}\left( x\right) &=&2x^{2}\left( 1-x\right) ,\hspace{1cm}\hspace{1cm}%
P_{A}\left( x\right) =2x^{2}\left( 1-x\right) \left( x-4\right) -4\frac{%
m_{\mu }^{2}}{M_{Z^{\prime }}^{2}}x^{3}  \notag \\
g_{L} &=&-3g_{X},\hspace{1cm}\hspace{1cm}g_{L}=-6g_{X},\hspace{1cm}\hspace{%
1cm}g_{V,A}=g_{R}\pm g_{L}\,.
\end{eqnarray}%
%
\begin{figure}[tbh]
\resizebox{18cm}{22cm}{\includegraphics{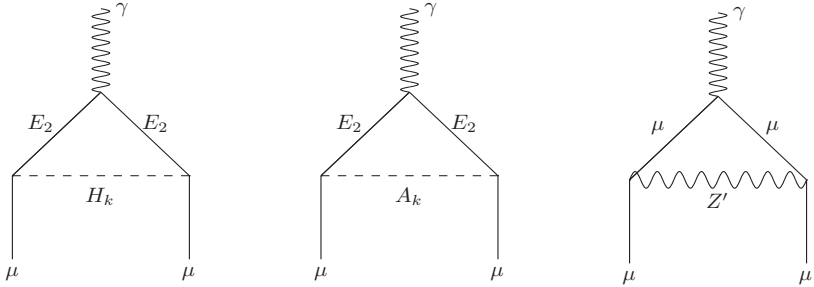}}\vspace{-16cm}
\caption{Loop Feynman diagrams contributing to the muon anomalous magnetic
moment. Here $k=1,2$}
\label{Loopdiagramsq}
\end{figure}

\begin{figure}[!h]
\resizebox{8.9cm}{6cm}{\includegraphics{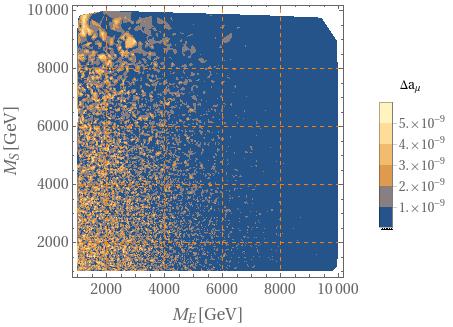}}\vspace{0cm} %
\resizebox{8.9cm}{6cm}{\includegraphics{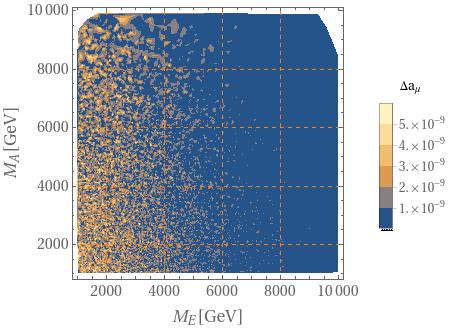}}\vspace{0cm}
\caption{Allowed parameter space for $M_{S}$-$M_{E}$ (left panel) and $M_{A}$%
-$M_{E}$ (right panel) planes with different values of the muon anomalous
magnetic moment.}
\label{Correlation}
\end{figure}

In our numerical analysis we have fixed $\tan \theta =\frac{v}{v_{\sigma }}$%
, $M_{Z^{\prime }}=1.5$ TeV and $g_{X}=0.1$, in consistency with the $%
2.6\sigma $ $R_{K}$ anomaly. Considering that the muon anomalous magnetic
moment is constrained to be in the range \cite%
{Hagiwara:2011af,Nomura:2018lsx,Nomura:2018vfz}, 
\begin{equation}
\left( \Delta a_{\mu }\right)_{\exp }=\left( 26.1\pm 8\right) \times
10^{-10} \,,
\end{equation}
we plot in Figure \ref{Correlation} the allowed parameter space for $M_{S}$-$%
M_{E}$ (left panel) and $M_{A}$-$M_{E}$ (right panel) planes with different
values for $\Delta a_{\mu }$. Here, we have set $M_{S}=\min
\left(m_{H_{1}},m_{H_{2}}\right) $ and $M_{A}=\min
\left(m_{A_{1}},m_{A_{2}}\right) $ and $M_{E}=m_{E_{2}}$. We found that our
model can accommodate the experimental values of $\Delta a_{\mu}$ for a
large region of parameter space.

\section{DM particle candidates}

\label{sec:DM} Note that due to the exact $Z_{2}^{(2)}$ discrete symmetry,
our model has several stable scalar DM (DM) candidates, which can be the
neutral components of the inert $SU_{2L}$ scalar doublet $\phi _{2}$ as well
as the real and imaginary parts of the SM scalar singlets $\sigma _{2}$, $%
\sigma _{3}$, $\rho _{1}$, $\rho _{2}$ and $\eta $. Furthermore, the model
can have a fermionic DM candidate, which is the only SM-singlet Majorana
neutrino $\Omega _{1R}$ with a non-trivial $Z_{2}^{(2)}$ charge. 

Considering a scenario with a scalar DM candidate, one has to ensure its
stability. 
This can be done by assuming that it is the lightest among the inert scalar
particles and is lighter than the exotic fermions. That scalar DM candidate
annihilates mainly into $WW$, $ZZ$, $t\overline{t}$, $b\overline{b}$ and $hh$
via a Higgs portal scalar interaction $\left(\phi^{\dagger}_1\phi_1\right)%
\Phi^{\dagger}_{DM}\Phi_{DM}$, where $\phi_1$ is the SM Higgs doublet and $%
\Phi_{DM}$ the scalar DM candidate in our model. These annihilation channels
will contribute to the DM relic density, which can be accommodated for
appropriate values of the scalar DM mass and of the quartic scalar coupling
of the corresponding Higgs portal scalar interaction $\left(\phi^{\dagger}_1%
\phi_1\right)\Phi^{\dagger}_{DM}\Phi_{DM}$, similarly as in Refs. \cite%
{CarcamoHernandez:2016pdu,Bernal:2017xat,CarcamoHernandez:2017kra,Long:2018dun}%
. Thus, for DM direct detection prospects, the scalar DM candidate would
scatter off a nuclear target in a detector via Higgs boson exchange in the $%
t $-channel, giving rising to a constraint on the coupling of the $%
\left(\phi^{\dagger}_1\phi_1\right)\Phi^{\dagger}_{DM}\Phi_{DM}$
interaction. Given the large number of parameters in the scalar potential of
our model (which is discussed in detail in our previous work \cite%
{CarcamoHernandez:2019cbd}), there is a lot of parametric freedom that
allows us to reproduce 

the observed value of the DM relic density and the parameter space of our
model consistent with DM constraints will be similar to the one in Refs.~%
\cite%
{CarcamoHernandez:2016pdu,Bernal:2017xat,CarcamoHernandez:2017kra,Long:2018dun}%
.

For instance, in the case, where $m_{{\Phi_{DM}}}^{2}>>v^{2}$, with $v=246$
GeV, and neglecting the annihilation channel of the scalar DM candidate into
neutrino-antineutrino pairs as in Ref.~\cite{Bernal:2017xat}, the freeze-out
of heavy scalar DM particle will be largely dominated by the annihilations
into Higgs bosons and the corresponding thermally averaged cross section can
be estimated as 
\begin{eqnarray}  \label{eq:sigma-1}
<\sigma v>\simeq \frac{\gamma^{2}}{128\pi m_{{\Phi_{DM} }}^{2}},
\end{eqnarray}
which results in a DM relic abundance 
\begin{eqnarray}  \label{eq:Omega-1}
\frac{\Omega _{DM}h^{2}}{0.12}=\frac{0.1pb}{0.12<\sigma v>}\simeq\left( 
\frac{1}{\gamma}\right) ^{2}\left( \frac{m_{{\Phi_{DM} }}}{1.1TeV}\right)
^{2},
\end{eqnarray}
with $\gamma$ being the quartic scalar coupling of the Higgs portal scalar
interaction $\left(\phi^{\dagger}_1\phi_1\right)\Phi^{\dagger}_{DM}\Phi_{DM}$%
. Consequently, our model naturally reproduces the observed value \cite%
{Ade:2015xua} 
\begin{eqnarray}  \label{eq:Omega-observed}
\Omega_{DM}h^{2}=0.1198
\end{eqnarray}
for the DM relic density.

In the scenario with a fermionic DM candidate, it follows from the Yukawa
interactions $\overline{\Omega _{1R}^{C}}\eta ^{\ast }\nu _{kR}$ and $%
\overline{\Omega _{1R}^{C}}\eta \Psi_{R}$ that the DM candidate $\Omega_{1R}$
can annihilate into a pair of the right-handed Majorana neutrinos $\nu_{kR}$
($k=1,3$) and $\Psi_R$, via $t$ channel exchange of the real and imaginary
parts of the gauge singlet scalar $\eta$. Additionally, the fermionic DM
candidate $\Omega _{1R}$ can also annihilate into $\eta_R\eta_R$ and $%
\eta_I\eta_I$ via the the $t$ channel exchange of the right handed Majorana
neutrinos $\nu_{kR}$ ($k=1,3$) and $\Psi_R$. 
Thus, the corresponding relic density will depend on the neutrino Yukawa
coupling of the aforementioned Yukawa interactions, on the fermionic DM
candidate mass $m_{\Omega_{1R}}$, on the masses of the the right-handed
Majorana neutrinos $\nu_{kR}$ ($k=1,3$), $\Psi_R$, as well as on the masses
of the real and imaginary parts of the gauge singlet scalar $\eta$.

Considering a scenario where $m_{{\Omega _{1R}}}^{2}{<<}m_{{\ \eta _{R}}%
}^{2}\sim m_{{\eta _{I}}}^{2}\sim m_{{\eta }}^{2}$, and the annihilation
channel $\Omega _{1R}\Omega _{1R}\rightarrow \nu _{kR}\nu _{kR}$ ($k=1,3$),
following Ref. \cite{Bernal:2017xat} one can estimate the corresponding
thermally averaged cross section as 
\begin{eqnarray}  \label{eq:sigma-2}
<\sigma v>\simeq \frac{9y_{\Omega }^{4}m_{{\Omega }}^{2}}{16\pi m_{{\eta }%
}^{4}}.
\end{eqnarray}

Then, the DM relic abundance is 
\begin{eqnarray}  \label{eq:Omega-2}
\frac{\Omega _{DM}h^{2}}{0.12}=\frac{0.1pb}{0.12<\sigma v>}\simeq \left( 
\frac{1}{y_{\Omega }}\right) ^{4}\left( \frac{400GeV}{m_{{\Omega }}} \right)
^{2}\left( \frac{m_{{\eta }}}{1.9TeV}\right) ^{4},
\end{eqnarray}
showing that in the case of fermionic DM candidate our model also naturally
reproduces the observed value (\ref{eq:Omega-observed}).



\section{Conclusions}


We have studied some phenomenological aspects of the extended Inert Higgs
Doublet model, which incorporates the mechanism of sequential
loop-generation of the SM fermion masses, 
explaining 
the observed strong hierarchies between them as well as the corresponding
mixing parameters. A particular emphasis has been made on analyzing the
constraints on the $Z^{\prime }$ mass and couplings of our model, imposed by
the $2.6\sigma $ anomaly in lepton universality, the LEP constraint on the $%
M_{Z^{\prime }}/g_{X}$ ratio and the constraints arising from the
experimental measurements of the muon anomalous magnetic moment.
Furthermore, we have studied production of the heavy $Z^{\prime }$ gauge
boson in proton-proton collisions via the Drell-Yan mechanism. We found that
the corresponding total cross section at the LHC is equal to $0.2-1$ pb when
the heavy $Z^{\prime }$ mass is varied within $1.5-2$ TeV interval for the $%
U_{1X}$ gauge coupling $g_X=0.1$. The $Z^{\prime }$ production cross section
gets significantly enhanced at a future $100$ TeV proton-proton collider
reaching the typical values of $9-29$ pb. Additionally, we have found that
the $2.6\sigma $ anomaly in lepton universality yields a tighter constraint
than the one obtained from the $e^{+}e^{-}\rightarrow \mu ^{+}\mu ^{-}$
measurement at LEP and implies a lower bound of $\approx 13$ TeV on the $%
M_{Z^{\prime }}/g_{X}$ ratio. 
We have found that our model successfully accommodates the experimental
values of the muon magnetic moment for a large region of parameter space.
Finally, we have examined the possible fermion and scalar DM particle
candidates of the model and showed that in both cases our predictions are
compatible with the observed DM relic density abundance.

\textbf{Acknowledgements} 
This research has received funding from Fondecyt (Chile) grants No.~1170803,
No.~1190845, No.~1180232, No.~3150472 and 
the UTFSM grant PIM175. R.P. is partially supported by the Swedish Research
Council, contract numbers 621-2013-4287 and 2016-05996, by CONICYT grant 
MEC80170112, as well as by the European Research Council (ERC) under the
European Union's Horizon 2020 research and innovation programme (grant
agreement No 668679). This work was supported in part by the Ministry of
Education, Youth and Sports of the Czech Republic, project LTC17018. A.E.C.H
thanks University of Lund, where part of this work was done, for hospitality
as well as University of Southampton and Institute of Experimental and
Applied Physics of the Czech Technical University in Praga for hospitality
during the completion of this work.


\begin{thebibliography}{99}

\bibitem{Campos:2014zaa} M.~D.~Campos, A.~E.~C\'arcamo Hern\'andez, H.~Pas
and E.~Schumacher, 
Phys.\ Rev.\ D \textbf{91}, no. 11, 116011 (2015)
doi:10.1103/PhysRevD.91.116011 [arXiv:1408.1652 [hep-ph]]. 


\bibitem{Hernandez:2015hrt} A.~E.~C\'arcamo Hern\'andez, 
Eur.\ Phys.\ J.\ C \textbf{76}, no. 9, 503 (2016)
doi:10.1140/epjc/s10052-016-4351-y [arXiv:1512.09092 [hep-ph]]. 


\bibitem{Hernandez:2015dga} A.~E.~C\'arcamo Hern\'andez, I.~de Medeiros
Varzielas and E.~Schumacher, 
Phys.\ Rev.\ D \textbf{93}, no. 1, 016003 (2016)
doi:10.1103/PhysRevD.93.016003 [arXiv:1509.02083 [hep-ph]]. 


\bibitem{Arbelaez:2016mhg} C.~Arbel\'aez, A.~E.~C\'arcamo Hern\'andez,
S.~Kovalenko and I.~Schmidt, 
Eur.\ Phys.\ J.\ C \textbf{77}, no. 6, 422 (2017)
doi:10.1140/epjc/s10052-017-4948-9 [arXiv:1602.03607 [hep-ph]]. 


\bibitem{Mantilla:2016lui} S.~F.~Mantilla, R.~Martinez and F.~Ochoa, 
Phys.\ Rev.\ D \textbf{95}, no. 9, 095037 (2017)
doi:10.1103/PhysRevD.95.095037 [arXiv:1612.02081 [hep-ph]]. 


\bibitem{Bernal:2017xat} N.~Bernal, A.~E.~C\'arcamo Hern\'andez, I.~de Medeiros Varzielas and S.~Kovalenko, 
JHEP \textbf{1805}, 053 (2018) doi:10.1007/JHEP05(2018)053 [arXiv:1712.02792
[hep-ph]]. 

\bibitem{CarcamoHernandez:2016pdu} A.~E.~C\'arcamo Hern\'andez, S.~Kovalenko
and I.~Schmidt, 
JHEP \textbf{1702}, 125 (2017) doi:10.1007/JHEP02(2017)125 [arXiv:1611.09797
[hep-ph]]. 

\bibitem{CarcamoHernandez:2017cwi} A.~E.~C\'arcamo Hern\'andez,
S.~Kovalenko, H.~N.~Long and I.~Schmidt, 
JHEP \textbf{1807}, 144 (2018) doi:10.1007/JHEP07(2018)144 [arXiv:1705.09169
[hep-ph]].



\bibitem{Mantilla:2017ijh} S.~F.~Mantilla and R.~Martinez, 
Phys.\ Rev.\ D \textbf{96}, no. 9, 095027 (2017)
doi:10.1103/PhysRevD.96.095027 [arXiv:1704.04869 [hep-ph]]. 



\bibitem{Abbas:2017vws} G.~Abbas, 
arXiv:1712.08052 [hep-ph]. 


\bibitem{Dev:2018pjn} A.~Dev and R.~N.~Mohapatra, 
Phys.\ Rev.\ D \textbf{98}, no. 7, 073002 (2018)
doi:10.1103/PhysRevD.98.073002 [arXiv:1804.01598 [hep-ph]]. 


\bibitem{CarcamoHernandez:2018hst} 
  A.~E.~C\'arcamo Hern\'andez, S.~Kovalenko, J.~W.~F.~Valle and C.~A.~Vaquera-Araujo,
  JHEP {\bf 1902}, 065 (2019)
  doi:10.1007/JHEP02(2019)065
  [arXiv:1811.03018 [hep-ph]].



\bibitem{Abbas:2018lga} G.~Abbas, 
arXiv:1807.05683 [hep-ph]. 



\bibitem{CarcamoHernandez:2019pmy} 
  A.~E.~C\'arcamo Hern\'andez, J.~Marchant Gonz\'alez and U.~J.~Saldaña-Salazar,
  arXiv:1904.09993 [hep-ph].


\bibitem{CarcamoHernandez:2019vih} 
  A.~E.~C\'arcamo Hern\'andez, Y.~Hidalgo Vel\'asquez and N.~A.~P\'erez-Julve,
  arXiv:1905.02323 [hep-ph].



\bibitem{CarcamoHernandez:2019cbd} A.~E.~C\'arcamo Hern\'andez,
S.~Kovalenko, R.~Pasechnik and I.~Schmidt, 
arXiv:1901.02764 [hep-ph]. 


\bibitem{Vicente:2018xbv} A.~Vicente, 
Adv.\ High Energy Phys.\ \textbf{2018}, 3905848 (2018)
doi:10.1155/2018/3905848 [arXiv:1803.04703 [hep-ph]]. 


\bibitem{Crivellin:2015era} A.~Crivellin, L.~Hofer, J.~Matias, U.~Nierste,
S.~Pokorski and J.~Rosiek, 
Phys.\ Rev.\ D \textbf{92} (2015) no.5, 054013
doi:10.1103/PhysRevD.92.054013 [arXiv:1504.07928 [hep-ph]]. 


\bibitem{Crivellin:2015lwa} A.~Crivellin, G.~D'Ambrosio and J.~Heeck, 
Phys.\ Rev.\ D \textbf{91} (2015) no.7, 075006
doi:10.1103/PhysRevD.91.075006 [arXiv:1503.03477 [hep-ph]]. 


\bibitem{King:2018fcg} S.~F.~King, 
JHEP \textbf{1809}, 069 (2018) doi:10.1007/JHEP09(2018)069 [arXiv:1806.06780
[hep-ph]]. 


\bibitem{Bonilla:2017lsq} C.~Bonilla, T.~Modak, R.~Srivastava and
J.~W.~F.~Valle, 
Phys.\ Rev.\ D \textbf{98}, no. 9, 095002 (2018)
doi:10.1103/PhysRevD.98.095002 [arXiv:1705.00915 [hep-ph]]. 


\bibitem{Barbieri:2017tuq} R.~Barbieri and A.~Tesi, 
Eur.\ Phys.\ J.\ C \textbf{78}, no. 3, 193 (2018)
doi:10.1140/epjc/s10052-018-5680-9 [arXiv:1712.06844 [hep-ph]]. 


\bibitem{King:2017anf} S.~F.~King, 
JHEP \textbf{1708}, 019 (2017) doi:10.1007/JHEP08(2017)019 [arXiv:1706.06100
[hep-ph]]. 


\bibitem{Romao:2017qnu} M.~C.~Romao, S.~F.~King and G.~K.~Leontaris, 
arXiv:1710.02349 [hep-ph]. 


\bibitem{Antusch:2017tud} S.~Antusch, C.~Hohl, S.~F.~King and V.~Susic, 
arXiv:1712.05366 [hep-ph]. 


\bibitem{Ko:2017quv} P.~Ko, T.~Nomura and H.~Okada, 
Phys.\ Lett.\ B \textbf{772}, 547 (2017) doi:10.1016/j.physletb.2017.07.021
[arXiv:1701.05788 [hep-ph]]. 


\bibitem{Ko:2017yrd} P.~Ko, T.~Nomura and H.~Okada, 
Phys.\ Rev.\ D \textbf{95}, no. 11, 111701 (2017)
doi:10.1103/PhysRevD.95.111701 [arXiv:1702.02699 [hep-ph]]. 


\bibitem{Chen:2017hir} C.~H.~Chen, T.~Nomura and H.~Okada, 
Phys.\ Lett.\ B \textbf{774}, 456 (2017) doi:10.1016/j.physletb.2017.10.005
[arXiv:1703.03251 [hep-ph]]. 


\bibitem{Assad:2017iib} 
  N.~Assad, B.~Fornal and B.~Grinstein,
  Phys.\ Lett.\ B {\bf 777}, 324 (2018)
  doi:10.1016/j.physletb.2017.12.042
  [arXiv:1708.06350 [hep-ph]].



\bibitem{Angelescu:2018tyl} A.~Angelescu, D.~Becirevic,
D.~A.~Faroughy and O.~Sumensari, 
JHEP \textbf{1810}, 183 (2018) doi:10.1007/JHEP10(2018)183 [arXiv:1808.08179
[hep-ph]]. 


\bibitem{DiLuzio:2018zxy} L.~Di Luzio, J.~Fuentes-Martin, A.~Greljo,
M.~Nardecchia and S.~Renner, 
JHEP \textbf{1811}, 081 (2018) doi:10.1007/JHEP11(2018)081 [arXiv:1808.00942
[hep-ph]]. 


\bibitem{Guadagnoli:2018ojc} D.~Guadagnoli, M.~Reboud and O.~Sumensari, 
JHEP \textbf{1811}, 163 (2018) doi:10.1007/JHEP11(2018)163 [arXiv:1807.03285
[hep-ph]]. 


\bibitem{Fornal:2018dqn} B.~Fornal, S.~A.~Gadam and B.~Grinstein, 
arXiv:1812.01603 [hep-ph]. 

\bibitem{Aydemir:2018cbb} 
  U.~Aydemir, D.~Minic, C.~Sun and T.~Takeuchi,
  JHEP {\bf 1809}, 117 (2018)
  doi:10.1007/JHEP09(2018)117
  [arXiv:1804.05844 [hep-ph]].


\bibitem{Faber:2018qon} T.~Faber, M.~Hudec, M.~Malinsky, P.~Meinzinger, W.~Porod and F.~Staub, 
Phys.\ Lett.\ B \textbf{787}, 159 (2018) doi:10.1016/j.physletb.2018.10.051
[arXiv:1808.05511 [hep-ph]]. 


\bibitem{Barman:2018jhz} B.~Barman, D.~Borah, L.~Mukherjee and S.~Nandi, 
arXiv:1808.06639 [hep-ph]. 


\bibitem{Heeck:2018ntp} J.~Heeck and D.~Teresi, 
JHEP \textbf{1812}, 103 (2018) doi:10.1007/JHEP12(2018)103 [arXiv:1808.07492
[hep-ph]]. 


\bibitem{Grinstein:2018fgb} B.~Grinstein, S.~Pokorski and G.~G.~Ross, 
JHEP \textbf{1812}, 079 (2018) [JHEP \textbf{2018}, 079 (2020)]
doi:10.1007/JHEP12(2018)079 [arXiv:1809.01766 [hep-ph]]. 


\bibitem{Falkowski:2018dsl} A.~Falkowski, S.~F.~King, E.~Perdomo and
M.~Pierre, 
JHEP \textbf{1808}, 061 (2018) doi:10.1007/JHEP08(2018)061 [arXiv:1803.04430
[hep-ph]]. 


\bibitem{CarcamoHernandez:2018aon} A.~E.~C\'arcamo Hern\'andez and
S.~F.~King, 
arXiv:1803.07367 [hep-ph]. 


\bibitem{deMedeirosVarzielas:2018bcy} I.~de Medeiros Varzielas and
S.~F.~King, 
JHEP \textbf{1811}, 100 (2018) doi:10.1007/JHEP11(2018)100 [arXiv:1807.06023
[hep-ph]]. 


\bibitem{Rocha-Moran:2018jzu} P.~Rocha-Moran and A.~Vicente, 
arXiv:1810.02135 [hep-ph]. 

\bibitem{Hu:2018veh} Q.~Y.~Hu, X.~Q.~Li and Y.~D.~Yang, 
arXiv:1810.04939 [hep-ph]. 

\bibitem{Carena:2018cow} M.~Carena, E.~Megias, M.~Quiros and C.~Wagner, 
JHEP \textbf{1812}, 043 (2018) doi:10.1007/JHEP12(2018)043 [arXiv:1809.01107
[hep-ph]]. 


\bibitem{Babu:2018vrl} K.~S.~Babu, R.~N.~Mohapatra and B.~Dutta, 
arXiv:1811.04496 [hep-ph]. 


\bibitem{Allanach:2018lvl} B.~C.~Allanach and J.~Davighi, 
arXiv:1809.01158 [hep-ph]. 

\bibitem{Hurth:2016fbr} T.~Hurth, F.~Mahmoudi and S.~Neshatpour, 
Nucl.\ Phys.\ B \textbf{909}, 737 (2016) doi:10.1016/j.nuclphysb.2016.05.022
[arXiv:1603.00865 [hep-ph]]. 


\bibitem{Glashow:1976nt} 
  S.~L.~Glashow and S.~Weinberg,
  Phys.\ Rev.\ D {\bf 15}, 1958 (1977).
  doi:10.1103/PhysRevD.15.1958



\bibitem{Paschos:1976ay} 
  E.~A.~Paschos,
  Phys.\ Rev.\ D {\bf 15}, 1966 (1977).
  doi:10.1103/PhysRevD.15.1966



\bibitem{Schael:2013ita} S.~Schael \textit{et al.} [ALEPH and DELPHI and L3
and OPAL and LEP Electroweak Collaborations], 
Phys.\ Rept.\ \textbf{532}, 119 (2013) doi:10.1016/j.physrep.2013.07.004
[arXiv:1302.3415 [hep-ex]]. 

\bibitem{Diaz:2002uk} R.~A.~Diaz, R.~Martinez and J.~A.~Rodriguez, 
Phys.\ Rev.\ D \textbf{67}, 075011 (2003) doi:10.1103/PhysRevD.67.075011
[hep-ph/0208117]. 

\bibitem{Kelso:2014qka} C.~Kelso, H.~N.~Long, R.~Martinez and F.~S.~Queiroz, 
Phys.\ Rev.\ D \textbf{90}, no. 11, 113011 (2014)
doi:10.1103/PhysRevD.90.113011 [arXiv:1408.6203 [hep-ph]]. 

\bibitem{Hagiwara:2011af} K.~Hagiwara, R.~Liao, A.~D.~Martin, D.~Nomura and
T.~Teubner, 
J.\ Phys.\ G \textbf{38}, 085003 (2011) doi:10.1088/0954-3899/38/8/085003
[arXiv:1105.3149 [hep-ph]]. 

\bibitem{Nomura:2018lsx} T.~Nomura and H.~Okada, 
arXiv:1808.05476 [hep-ph]. 

\bibitem{Nomura:2018vfz} T.~Nomura and H.~Okada, 
Phys.\ Rev.\ D \textbf{97}, no. 9, 095023 (2018)
doi:10.1103/PhysRevD.97.095023 [arXiv:1803.04795 [hep-ph]]. 


\bibitem{CarcamoHernandez:2017kra} A.~E.C\'arcamo Hern\'andez and
H.~N.~Long, 
J.\ Phys.\ G \textbf{45}, no. 4, 045001 (2018) doi:10.1088/1361-6471/aaace7
[arXiv:1705.05246 [hep-ph]]. 


\bibitem{Long:2018dun} H.~N.~Long, N.~V.~Hop, L.~T.~Hue, N.~H.~Thao and
A.~E.~C\'arcamo Hern\'andez, 
arXiv:1810.00605 [hep-ph]. 


\bibitem{Ade:2015xua} P.~A.~R.~Ade \textit{et al.} [Planck Collaboration], 
\emph{\ ``Planck 2015 results. XIII. Cosmological parameters,'' Astron.\
Astrophys.}\ \textbf{594}, A13 (2016) 
[arXiv:1502.01589 [astro-ph.CO]]. 
\end{thebibliography}
\end{document}